\documentclass{article}

\usepackage{graphicx}
\usepackage{amssymb}
\usepackage{natbib}
\usepackage{graphicx}
\usepackage{latexsym}
\usepackage{ marvosym }
\usepackage{algorithmic}
\usepackage[ruled,vlined,linesnumbered]{algorithm2e}
\usepackage{amsfonts}
\usepackage{booktabs}
\usepackage{siunitx}
\usepackage{multirow}
\usepackage{rotating}
\usepackage{hhline}
\usepackage{dsfont}
\usepackage{amsmath}
\usepackage{enumerate}

\newtheorem{corollary}{Corollary}
\newtheorem{example}{Example}
\newtheorem{theorem}{Theorem}
\newcommand{\qed}{\hfill $\square$}
\newcommand{\proof}{\noindent \emph{Proof. }}

\newcommand{\items}{\mathcal{I}} 			

\newcommand{\R}[2]{{#1}\rightarrow {#2}}

\newcommand{\SDB}{\mathcal{D}}	

\SetKw{myproc}{procedure}
\SetKw{myfunc}{function}
\SetKw{Inp}{in}
\SetKw{Outp}{out}
\SetKw{InOutp}{in out}
\SetKw{F}{false}
\SetKw{T}{true}
\SetKw{lAnd}{and}
\SetKw{Comment}{$\triangleright$\ }{}

\usepackage{etoolbox}
\newbool{seeAll}
\usepackage[usenames,dvipsnames]{color}
\usepackage[normalem]{ulem}

\usepackage{xcolor}
\usepackage{soul}
\usepackage[utf8]{inputenc}
\usepackage[small]{caption}

\newcommand{\conf}[0]{\ensuremath{conf}\xspace}

\newcommand{\body}{{body}\xspace}
\newcommand{\head}{{head}\xspace}

\newcommand{\minConf}{{\sc c}}

\makeatletter
\newcommand{\verbatimfont}[1]{\renewcommand{\verbatim@font}{\ttfamily#1}}
\makeatother

\title{ 
Computational Complexity of Three Central Problems  in Itemset Mining
\thanks{This work was partially supported by the T-LARGO project.}
}

\author{
	Christian Bessiere\\CNRS, University of Montpellier, France\and 
	Mohamed-Bachir Belaid\\ Simula Research Laboratory, Lysaker, Norway\and
	Nadjib Lazaar\\LIRMM, University of Montpellier, CNRS, Montpellier, France 
}

\begin{document} 



\maketitle


\begin{abstract}
Itemset mining is one of the most studied tasks in knowledge discovery.
In this paper we analyze the computational complexity of three central 
itemset mining problems. 
We  prove that mining confident rules 
with a given item in the head is NP-hard. 
We prove that  mining  high utility itemsets is NP-hard. 
We finally prove that mining maximal or closed itemsets  is coNP-hard as soon 
as the users can specify constraints on the kind of itemsets they are interested in. 
\end{abstract}

\section{Introduction}
Many techniques have been developed for itemset mining problems. 
Famous examples are  mining frequent itemsets \citep{agrawal1993mining,han2000mining}, 
mining association rules \citep{agrawal1993mining,szathmary2007zart}, 
mining closed itemsets \citep{pasquier1999discovering}, 
mining high utility itemsets \citep{chan2003mining}, etc.
Most of these works have focused on improving the practical performance of the mining process, but few have conducted a theoretical analysis of the computational complexity 
of itemset mining problems. 

\cite{DBLP:journals/datamine/WijsenM98} have proved 
that it is NP-complete to decide whether there exists a valid quantitative rule. 
\cite{DBLP:conf/sebd/AngiulliIP01} have proved that the  problem of deciding 
whether there exists a non-redundant association 
rule of size at least $k$ that is frequent and 
the problem of deciding 
whether there exists a non-redundant association 
rule of size at least $k$ that is confident are 
both NP-complete. 
There also exist results on  the computational complexity  
of mining maximal frequent itemsets. 
\cite{DBLP:conf/kdd/Yang04}, and \cite{zaki1998theoretical} have proved that 
deciding whether there exists a maximal frequent itemset is polynomial. 
\cite{DBLP:conf/stacs/BorosGKM02} have proved that deciding 
whether  there exist other  maximal frequent itemsets 
than those in a given set is NP-complete.

In this paper we analyze the computational complexity of three well-known itemset mining problems. 
We prove that  deciding whether there exists a confident rule that contains a given item in the head is NP-complete. 
This directly leads to the result that mining confident rules 
with a given item in the head is NP-hard. 
We then prove that  deciding whether there exists an itemset with high utility is NP-complete.
This directly leads to the result that  
mining  high utility itemsets is NP-hard. 
We finally prove that  deciding whether 
there exists an itemset that is maximal or closed
w.r.t.  those  
satisfying a set of user's constraints is coNP-complete.
This directly leads to the result that mining maximal or closed  
constrained itemsets   is coNP-hard.

The paper is organized as follows. We start with some preliminary 
definitions and notations in Section \ref{sec:notations}.
In Section \ref{sec:conf} we study the problem of mining association rules that are confident.
Section \ref{sec:high} reports our result on the problem of mining high utility itemsets.
In Section \ref{sec:borders} we study the problem of mining maximal or closed
itemsets among itemsets subject to a set of user's constraints. 
Section \ref{sec:conclusion}  concludes this work.

\section{Preliminary Definitions and Notations}
\label{sec:notations}

Let $\items = {p_1,\ldots,p_n}$ be a set of $n$ distinct objects, called \emph{items}. 
An \emph{itemset} $P$ is a non-empty subset of $\items$. 
A \emph{transactional dataset} $\SDB$ is a bag of $m$ itemsets 
$t_1,\ldots,t_m$, called \emph{transactions}. 

The \emph{cover} of an itemset $P$ in $\SDB$, denoted by $cover(P)$, 
is the bag of transactions from $\SDB$ containing $P$.
The \emph{frequency} of an itemset $P$ in $\SDB$, denoted by $freq(P)$, 
is the cardinality of its cover, i.e. $freq(P)=|cover(P)|$. 
Given a  frequency threshold $s$, an itemset $P$ is \emph{frequent} in $\SDB$ 
if $freq(P) \geq s$.  
This condition  is called  the minimum frequency 
constraint.

\begin{example}\label{exa:dataset}
\begin{table}[b]
	\caption{ Dataset  with five items and five transactions.}\label{Tab:example}
		\centering
		
		\begin{tabular}{c|c*4{@{\ }c@{\ }}}
			\hline
			trans. & \multicolumn{5}{c}{Items}        \\ \hline
			$t_1$ & $A$ & $B$ & $ $ & $D$ & $E$\\ 
			$t_2$ & $A$ & $ $ & $C$ & $ $ & $ $\\ 
			$t_3$ & $A$ & $B$ & $C$ & $ $ & $E$ \\ 
			$t_4$ & $ $ & $B$ & $C$ & $ $ & $E$\\
			$t_5$ & $A$ & $B$ & $C$ & $ $ & $E$ \\ \hline
				
		\end{tabular}
\end{table}
The dataset in Table \ref{Tab:example} has 5 items and 5 transactions.
The cover of $CE$ is $cover(CE) = \{t_3, t_4, t_5\}$. 
Its frequency is the cardinality of its cover, i.e. $freq(CE) = |cover(CE)| = 3$. 
If the frequency threshold $s$ is equal to $2$, 
$CE$ is frequent. 
\end{example}

\section{On Mining Confident Rules }
\label{sec:conf}

\subsection{Background on association rules}

An {\em association rule} \citep{agrawal1993mining} is an implication of the 
form $\R{X}{Y}$, where $X$ and $Y$ are itemsets such that 
$X\cap Y=\emptyset$ and $Y \neq \emptyset$. 
$X$ represents the {\em \body{}} of the rule and $Y$ represents its {\em \head}.
The {\em confidence} of a rule captures how often $Y$ occurs in transactions 
containing $X$, that is, 
$\conf(\R{X}{Y})=\frac{freq({X}\cup{Y})}{freq(X)}$.  
Given  a confidence threshold $\minConf$, 
a rule  $\R{X}{Y}$  is confident if  $\conf(\R{X}{Y})\geq \minConf$.

\begin{example}
Consider the  dataset presented in Table \ref{Tab:example} and the confidence threshold $\minConf=60\%$.  
$B \to C$ is a 
confident association rule because 
$conf(B \to C) = \frac{freq(\{B, C\})}{freq(\{B\})} = \frac{3}{4}\geq c$. 
\end{example}

\subsection{Our result}

In this subsection we analyze the computational complexity 
of  mining  confident rules. 
We prove that deciding whether  there exists a confident rule with a given  item in the head is NP-complete, which implies that mining  confident rules that contain a given item in the head is NP-hard.

\begin{theorem}\label{th:confidence-completeness}
Given a dataset $\SDB$ on a set of items $\items$, 
deciding  whether there exists a rule that contains a
given item in the head and has a 
confidence higher than a given threshold $c$ is 
NP-complete.
\end{theorem}

\proof

\emph{Membership.}
Given an association rule $X\to Y$, we check if the 
given item belongs to $Y$. This 
is linear in the size of the rule. 
We then traverse the dataset $\SDB$ and compute  
the size of the covers of the itemsets $X$ and  $X\cup Y$. 
This is linear in $|\SDB|$. 
We then  compute the ratio $\frac{|cover(X\cup Y)|}{|cover(X)|}$ and 
compare it to $\minConf$  to decide if the rule is confident. This is linear in $log|\SDB|$. \\
\emph{Completeness.}
We reduce 3-SAT to the problem of deciding whether there exists a 
confident rule $X\to Y$ with a given  item in the head. 
Let us call \texttt{z} that item. 
Given a 3-SAT formula $F$ with $m$ 3-clauses 
on the set  $V = \{v_1,\ldots, v_n\}$ of 
Boolean variables, we build the following instance. 
For clarity purpose, we denote the items in the 
dataset $\SDB$  by \texttt{pos1}, \texttt{neg1}, $\cdots$, \texttt{posn}, \texttt{negn}, and \texttt{z}. 
We denote by \texttt{All} the set of all items. 
The confidence ratio $\minConf$ is set to $0.5$. 

The dataset $\SDB$ is:

\verbatimfont{\small}
\begin{enumerate}
\itemsep0ex
\item \label{trans:t1}  $ \texttt{All}\setminus \{\texttt{z}\} $\hfill  ($n$ times)
\item \label{trans:t2} $\texttt{All}\setminus\{\texttt{posi} \} , \forall v_i\in V$
\item \label{trans:t3} $\texttt{All}\setminus\{\texttt{negi} \} ,\forall v_i\in V$
\item \label{trans:t4} $\texttt{All}\setminus\{\texttt{posi},\texttt{negi},\texttt{z}\}, 
\forall v_i\in V$ \hfill (2 times)
\item \label{trans:t5} $\texttt{All}\setminus\{{it_1},{it_2} ,{it_3},\texttt{z}\}$, for each clause $cl$ in $F$, 
where $it_i = \texttt{posj}$ if the $i$th literal in $cl$ is $v_j$, 
$it_i = \texttt{negj}$ if the $i$th literal in $cl$ is $\neg v_j$. 
\end{enumerate}
The dataset contains $n+n+n+2n+m$ transactions. 
The reduction is thus polynomial in size. 

For instance, if $F=\{v_1 \lor \neg v_2\lor v_3\}$ with $n = 3$, $\SDB$ is: 
\begin{verbatim}
t1  pos1 |neg1 |pos2 |neg2 |pos3 |neg3 |   |
t2  pos1 |neg1 |pos2 |neg2 |pos3 |neg3 |   |
t3  pos1 |neg1 |pos2 |neg2 |pos3 |neg3 |   |
t4       |neg1 |pos2 |neg2 |pos3 |neg3 | z |
t5  pos1 |     |pos2 |neg2 |pos3 |neg3 | z |
t6       |     |pos2 |neg2 |pos3 |neg3 |   |
t7       |     |pos2 |neg2 |pos3 |neg3 |   |
t8  pos1 |neg1 |     |neg2 |pos3 |neg3 | z |
t9  pos1 |neg1 |pos2 |     |pos3 |neg3 | z |
t10 pos1 |neg1 |     |     |pos3 |neg3 |   |
t11 pos1 |neg1 |     |     |pos3 |neg3 |   |
t12 pos1 |neg1 |pos2 |neg2 |     |neg3 | z |
t13 pos1 |neg1 |pos2 |neg2 |pos3 |     | z |
t14 pos1 |neg1 |pos2 |neg2 |     |     |   |
t15 pos1 |neg1 |pos2 |neg2 |     |     |   |
t16      |neg1 |pos2 |     |     |neg3 |   |
\end{verbatim}

Suppose a formula $F$ is satisfiable. 
Let us denote by $S$ a solution of $F$. 
We construct the rule $X\to \{\texttt{z}\}$ such that  
$\texttt{posi}\in X$ and $\texttt{negi} \not\in X$ for each $i$ such that $S[v_i]=1$, and
$\texttt{posi}\not\in X$ and $\texttt{negi} \in X$ for each $i$ such that $S[v_i]=0$. 
By  construction of $\SDB$, 
$X\to \{\texttt{z}\}$ appears in $n$ transactions (\ref{trans:t2}) and (\ref{trans:t3}).  
By construction again, $X$ appears in the $n$ transactions 
where $X\to \{\texttt{z}\}$ appears 
plus the $n$ transactions   (\ref{trans:t1}). 
$X$ does not appear in any  transaction  (\ref{trans:t4}) 
because they all miss \texttt{posi}  and  \texttt{negi} for some $i$, 
whereas $X$ contains \texttt{posi}  or  \texttt{negi} for all $i$. 
Finally, as $S$ satisfies $F$, 
$X$ does not appear in any transaction (\ref{trans:t5}) because 
these transactions all miss at least the item of $X$ corresponding 
to the literal satisfying the clause. 
As a result, the rule $X\to \{\texttt{z}\}$ has confidence $\frac{n}{2n}=0.5 \geq c$.

Suppose now that $X\to Y$ is a confident rule with $\{\texttt{z}\}\in Y$. 
$Y$  contains \texttt{z}, so $X$ does not. 
Hence, $X$ appears at least in the $n$ transactions  (\ref{trans:t1})
where $Y$ does not appear. 
Now, $Y$  only appears in transactions (\ref{trans:t2}) and (\ref{trans:t3}) 
because 
it contains $\texttt{z}$.  
Thus,  $X$ must appear in at least $n$ transactions  (\ref{trans:t2}) and (\ref{trans:t3}) 
to reach the confidence of $50\%$. 
For a given $i$,  
$X$ must contain at least one among $\texttt{posi}$ and $\texttt{negi}$, 
otherwise the two corresponding transactions  (\ref{trans:t4}) would cover 
$X$ and not $Y$, making confidence impossible to reach. 
Thus, $X$ can (and must) appear in exactly $n$ transactions (\ref{trans:t2})  and (\ref{trans:t3}), 
which means that for each $i$, exactly one among $\texttt{posi}$ and $\texttt{negi}$ 
is in $X$. 
We then can build the mapping from the rule to 
the instantiation $S$ on $v_1,\ldots, v_n$ such that 
$S[v_i]=1$ if $\texttt{posi}\in X$, and $S[v_i]=0$ if $\texttt{negi}\in X$. 
We have $n$ transactions  (\ref{trans:t1}-\ref{trans:t4}) covering $X\cup Y$  
and $2n$ covering $X$.  
As  transactions   (\ref{trans:t5})  do not contain $\texttt{z}$, they must not 
cover  $X$, otherwise confidence cannot be reached. 
As a result, for every transaction  (\ref{trans:t5}), 
$X$  necessarily contains at least one item  (other than \texttt{z}) 
which is not in the transaction. 
By construction of transactions  (\ref{trans:t5})  and thanks to the mapping from the rule to $S$, 
this item corresponds 
to the truth value of a Boolean variable that satisfies the  clause of $F$ 
associated with the transaction. Therefore, $F$ is satisfiable. 

Consequently, 
deciding whether there exists a confident association rule 
with a given item in the head is NP-complete. 
\qed

\begin{corollary}\label{th:confidence-hardness2}
Given a dataset $\SDB$ on a set of items $\items$, 
finding a rule containing a given item in the head and having 
confidence higher than a given threshold $c$ is NP-Hard.
\end{corollary}

\section{On Mining High Utility Itemsets }
\label{sec:high}

\subsection{Background on high utility itemset mining}

In high utility itemset mining  \citep{chan2003mining}, each 
transaction $t_j$ is associated with a vector  $v_j$ of $n$ positive integers, 
where $v_j(i)$ is the cardinality of item $p_i$ in 
transaction $t_j$. 
A utility function $u$ is a vector of $n$ positive integers, 
where $u(i)$ is the utility of item $p_i$. 
The utility can be seen as the profit obtained 
when someone buys item $p_i$. 
The utility $u(P,t_j)$ of an itemset $P$ 
in a transaction  $t_j$ is 0 if $P\nsubseteq t_j$, 
$\sum_{p_i\in P}(v_j(i)\cdot u(i))$ otherwise. 
The utility $u(P)$ of an itemset $P$ 
is $\sum_{t_j\in cover(P)}u(P,t_j)$. 
Given a utility threshold $ut$, 
the itemset $P$ is of high utility if and only if $u(P)\geq ut$.

\begin{example}\label{exa: dataset-hui}
In Table \ref{Tab:value} we present the vectors $v_j$ for every 
transaction $t_j$ and the utility function $u(i)$ for every item.
The utility $u(AC,t_1)$ 
of the itemset $AC$ in $t_1$ is  0 because $AC\nsubseteq t_1$.
The utility  $u(AC,t_2)$ of the itemset $AC$ in $t_2$ is 
$u(AC,t_2) = 4\times 25 + 8\times 12 = 196$.
The utility  $u(AC)$ of the itemset $AC$ is 
$u(AC) = u(AC,t_2) + u(AC,t_3) + u(AC,t_5) = 196 + 218 + 282 = 696$.
If the utility threshold  $ut$ is set to $660$, $AC$ is of high 
utility ($u(AC)\geq ut$), but $ACE$ is not ($u(ACE) = 656 < ut$).

\begin{table}[h]
	\caption{  Dataset (left) with cardinality vectors (middle) 
	and utility function  (right).}\label{Tab:value}
		\centering
				\begin{tabular}{c|c*4{@{\ }c@{\ }}}
					\hline
					trans. & \multicolumn{5}{c}{Items}        \\ \hline
					$t_1$ & $A$ & $B$& $ $ & $D$ & $E$\\ 
					$t_2$ & $A$ &  & $C$ &  & \\ 
					$t_3$ & $A$ & $B$ & $C$ & & $E$ \\ 
					$t_4$ & & $B$ & $C$ & $ $ & $E$\\
					$t_5$ & $A$ & $B$ & $C$ & & $E$ \\ 
						
				\end{tabular}
				\quad\quad\quad
		\begin{tabular}{c|c*4{@{\ }c@{\ }}}
			\hline
			$v_j$ & $A$ & $B$ & $C$ & $D$ & $E$\\ \hline
			$v_1$ & 5 & 7 & 0 & 3 & 1\\ 
			$v_2$ & 4 & 0 & 8 & 0 & 0\\ 
			$v_3$ & 2 & 11 & 14 & 0 & 3 \\ 
			$v_4$ & 0 & 9 & 24 & 0 & 1\\
			$v_5$ & 6 & 5 & 11 & 0 & 2 \\
		\end{tabular}
		\quad\quad\quad
		\begin{tabular}{c|c*4{@{\ }c@{\ }}}
			\hline
			 & $A$ & $B$ & $C$ & $D$ & $E$\\ \hline
			$u$ & 25 & 14 & 12 & 36 & 34 \\
				
		\end{tabular}
\end{table}
\end{example}

\subsection{Our result}

In this subsection we analyze the computational 
complexity of  mining itemsets of high utility.  
Several algorithms have been proposed for mining high utility itemsets 
\citep{liu2005two,fournier2014fhm,peng2017mhuiminer,duong2018efficient}. 
None of these algorithms has been proved to be polynomial in time. 
We prove that 
deciding 
whether there exists a high utility itemset is NP-complete, 
which implies that mining  high utility itemsets is NP-hard. 
It is thus not possible that an algorithm for mining high 
utility itemsets is polynomial, unless $P=NP$.

\begin{theorem}
Given a dataset $\SDB$ on a set of items $\items$, 
given a utility function $u$, 
deciding whether there exists an itemset with 
utility higher than a given threshold $ut$ is 
NP-complete. 
\end{theorem}

\proof
\emph{Membership. }
Checking that an itemset $P$ is a witness to the 
existence of itemsets with utility higher than the 
threshold is done by computing the utility $u(P,t_j)$ of $P$ 
in each transaction $t_j$ such that $j$ is in the cover of $P$ and to sum these utilities. All this is  polynomial 
in $|\SDB|$.

\noindent
\emph{Completeness. }
We reduce \textsc{1in3-Positive-3SAT}, which is NP-complete,  
to the problem of deciding whether there exists 
an itemset with utility higher than the threshold $ut$. 
Given a formula $F$ with $m$ positive 
3-clauses on $n$ Boolean variables $v_1,\ldots,v_n$, 
we want to know whether there exists an assignment 
of the variables such that exactly one variable is 
true in each 3-clause. 
We build the dataset $\SDB$ on the set $\items=(p_1, \ldots, p_n)$  of items, 
where the  item $p_i$ represents the Boolean variable  $v_i$. 
An itemset $P$ corresponds to the  
assignment of the variables of formula $F$ such that 
$p_i\in P$ if and only if   $v_i=1$. 
The  utility function $u$ returns 1 for every item. 
$\SDB$ contains $3 m$ transactions and  
the utility threshold $ut$ is set to $3 nm^2$. 
Each clause $v_i\lor v_j\lor v_k$ in $F$ 
is encoded by adding three transactions to the dataset $\SDB$. 
These three transactions  have utility 1 
for all items except $p_i, p_j, p_k$. 
The first of the three transactions has utility $3nm$ for 
$p_i$, the second transaction has utility $3nm$ for $p_j$, 
and the third transaction has utility $3nm$ for $p_k$. 
The remaining two unset items in each transaction have utility 0. 
The dataset contains $3m$ transactions. 
The reduction is thus polynomial in size.

For instance, if 
$F = (v_1\vee v_2\vee v_3)\wedge (v_2\vee v_4\vee v_5)$ 
with $n = 5$ and $m = 2$, then $ut = 3\times 5\times 2^2=60$. 
The dataset $\SDB$ and the corresponding utilities and cardinalities (the vectors $v_i$) are presented in Table \ref{Tab:high}. 

\begin{table}[h]
	\caption{\small Example of an instance with $F = (v_1\vee v_2\vee v_3)\wedge (v_2\vee v_4\vee v_5)$, $m = 2$ and $n = 5$.}\label{Tab:high}
		\centering
		\begin{tabular}{c|c*4{@{\ }c@{\ }}}
			\hline
			trans. & \multicolumn{5}{c}{Items}        \\ \hline
			$t_1$ & $p_1$ & $ $ & $ $ & $p_4$ & $p_5$\\ 
			$t_2$ & $ $ & $p_2$ & $ $ & $p_4$ & $p_5$\\ 
			$t_3$ & $ $ & $ $ & $p_3$ & $p_4$ & $p_5$ \\ 
			$t_4$ & $p_1$ & $p_2$ & $p_3$ & $ $ & $ $\\
			$t_5$ & $p_1$ & $ $ & $p_3$ & $p_4$ & $ $ \\
			$t_6$ & $p_1$ & $ $ & $p_3$ & $ $ & $p_5$ \\ \hline
				
		\end{tabular}
		\quad\quad
		\begin{tabular}{c|c*4{@{\ }c@{\ }}}
			\hline
			$v_j$ & $p_1$ & $p_2$ & $p_3$ & $p_4$ & $p_5$\\ \hline
			$v_1$ & 30 & 0 & 0 & 1 & 1\\ 
			$v_2$ & 0 & 30 & 0 & 1 & 1\\ 
			$v_3$ & 0 & 0 & 30 & 1 & 1 \\ 
			$v_4$ & 1 & 30 & 1 & 0 & 0\\
			$v_5$ & 1 & 0 & 1 & 30 & 0 \\
			$v_6$ & 1 & 0 & 1 & 0 & 30 \\ 
				
		\end{tabular}\bigskip
		\quad\quad
		\begin{tabular}{c|c*4{@{\ }c@{\ }}}
			\hline
			 & $p_1$ & $p_2$ & $p_3$ & $p_4$ & $p_5$\\ \hline
			$u$ & 1 & 1 & 1 & 1 & 1 \\ 
		\end{tabular}
\end{table}

We show that there is a one-to-one mapping between 
itemsets with utility higher than the threshold and solutions 
to the  \textsc{1in3-Positive-3SAT} problem. 
Suppose that an itemset $P$ has utility higher than the threshold 
$3nm^2$. To reach this threshold, at least $m$ of the occurrences 
of $3nm$ in $\SDB$ must participate to the sum because the sum of all 
other occurrences in $\SDB$ does not reach $3nm$ (it is equal to $3(n-3)m$).
By construction of $\SDB$, if an itemset $P$ contains 
more than one item corresponding to variables in  
a given clause in $F$, none of the occurrences 
of $3nm$ in the three transactions encoding that clause 
can participate to the sum. 
As a result, to reach the threshold, 
for each triplet of transactions encoding a clause, 
an itemset $P$ must contain  exactly one of the three items corresponding to the 
three variables of the clause encoded by this triplet of transactions. 
By construction of these triplets of clauses, 
this means that assigning the set of Boolean variables corresponding to $P$ to value 1
and the others to 0, we obtain a solution to $F$. 

Suppose now that an assignment $A$ of the $v_i$'s is solution to  $F$. 
By definition, for each clause $v_i\lor v_j\lor v_k$, 
the itemset $P$  corresponding to $A$ contains 
exactly one item among $p_i, p_j, p_k$. 
By construction of the triplets of transactions representing the clauses, 
the sum of utilities contains an occurrence of $3nm$ per triplet, 
and thus  reaches the threshold. 

Consequently, 
deciding whether there exists an itemset with 
utility higher than a given threshold $ut$ is 
NP-complete.
\qed

\begin{corollary}
Given a dataset $\SDB$ on a set of items $\items$, 
given a utility function $u$, 
finding an itemset  with 
utility higher than a given threshold $ut$ is 
NP-hard. 
\end{corollary}

\section{On Mining Maximal or Closed Itemsets}\label{sec:borders}

\subsection{Background on constrained theories and concise representations}

Following \cite{bonchi2004closed}, 
a \emph{constraint} on itemsets is a function $c : 2^\items \to \{true,false\}$. 
We say that an itemset $P$ satisfies a constraint $c$ if and only if $c(P) = true$. 
Given a set $C$ of constraints and a dataset $\SDB$, 
the \emph{theory} of $C$  is the set of itemsets satisfying the constraints in $C$:
$$Th_{\SDB}(C) = \{P\in 2^\items\mid \forall c\in C: c(P) \}$$
Constraints are used to specify the kind of properties the user wants the 
mined itemsets to satisfy. 
For instance, given a  frequency threshold $s$, 
the minimum frequency constraint defined in Section \ref{sec:notations} 
is  denoted by $c_{freq}$: 
$$c_{freq}(P)\iff freq(P)\geq s$$
The theory $Th_{\SDB}(\{c_{freq}\})$ corresponds to the 
set of frequent itemsets.
Users may define any kind of constraints so that $Th_{\SDB}(C)$ corresponds 
to the itemsets they are interested in.

We now define two types of concise representations  
for a theory. 
The first one is  defined w.r.t. inclusion. 
An itemset  $P$ is {\emph{maximal}} for a theory 
if and only if $P$ is in the theory and 
none of its supersets are in the theory, that is, 
$$P\ \textrm{is maximal for } Th_{\SDB}(C) \iff 
P\in Th_{\SDB}(C) \land\nexists Q \in Th_{\SDB}(C) \mid Q \supsetneq  P$$

We can observe that when restricting our attention to 
the frequency constraint $c_{freq}$, 
itemsets that are maximal for $Th_{\SDB}(\{c_{freq}\})$ 
correspond to {maximal frequent itemsets} ({MFIs}) 
\citep{mannila1997levelwise}. 
%

There exists a  more restrictive type of concise representations 
that, in addition to inclusion, 
take into account the exact frequency of itemsets. 
%
An itemset  $P$ is {\emph{closed}} for a theory 
if and only if $P$ is in the theory and $P$ 
does not have any superset in the theory with the same frequency, that is, 
$$P\ \textrm{is closed for } Th_{\SDB}(C) \iff$$
$$P \in Th_{\SDB}(C) \land 
\nexists Q \in Th_{\SDB}(C) \mid Q\supsetneq P\land freq(Q)=freq(P)$$
%
%


\begin{example}\label{exa: dataset-concise}
In Example \ref{exa:dataset}, we saw that the itemset $CE$ is frequent 
in the dataset described in  Table \ref{Tab:example}. 
That is, 
$CE\in Th_{\SDB}(\{ c_{freq}\})$. $CE$ is not closed for   
$Th_{\SDB}(\{ c_{freq}\})$ because $BCE$ belongs to $Th_{\SDB}(\{ c_{freq}\})$ and 
$freq(BCE)=freq(CE)$. $BCE$ is closed for $Th_{\SDB}(\{ c_{freq}\})$ 
because it belongs to $Th_{\SDB}(\{ c_{freq}\})$ and 
none of its supersets have the  same frequency. 
$BCE$ is not maximal for 
$Th_{\SDB}(\{c_{freq}\})$ because $ABCE$ is a superset of $BCE$ and it also belongs to 
$Th_{\SDB}(\{c_{freq}\})$. 
\end{example}

Users often want to mine  
concise representations of a theory $Th_{\SDB}(C)$, 
where $C$ is a set of constraints, usually containing the frequency constraint 
$c_{freq}$, but also other constraints specifying  properties 
the returned itemsets should satisfy. 
There exist extremely efficient algorithms for mining concise representations 
of the  theory of frequent itemsets  $Th_{\SDB}(\{c_{freq})\}$. 
We can cite  the algorithm  CHARM for mining maximal 
frequent itemsets \citep{zaki2002charm},  or 
LCM for mining closed frequent itemsets \citep{uno2004lcm}. 
However, as noticed by \cite{bonchi2004closed}, 
given a set of constraints $C$ containing $c_{freq}$, 
mining itemsets that are maximal or closed
for the theory $Th_{\SDB}(C)$ does not simply consist in generating the 
maximal/closed
frequent itemsets and 
then remove those that do not satisfy the other constraints in $C$. 
It consists in mining those itemsets that are 
maximal/closed
for $Th_{\SDB}(C)$. 

Existing approaches for mining 
itemsets that are maximal/closed for a theory, such as the one presented in 
\citep{negrevergne2013dominance},
are ”multi-shot”, in the sense that they perform several calls to 
a SAT or CSP\footnote{The Constraint Satisfaction Problem (CSP) is a 
powerful paradigm to solve combinatorial problems \citep{rossi2006handbook}.} solver. 
The result in the next subsection shows that it is not possible to 
do differently (that is, "one-shot")  
unless 
$coNP\subseteq NP$. 

\subsection{Our result}

In this subsection we prove that it is coNP-hard to find itemsets that are 
maximal or closed for a theory  $Th_{\SDB}(C)$.

\begin{theorem}\label{th:max-coNP}
Given a dataset $\SDB$ on a set of items $\items$ and a set $C$ of  constraints, 
deciding whether an itemset  is maximal/closed 
for   $Th_{\SDB}(C)$
is coNP-complete, and it remains so even if $c_{freq}\in C$. 
\end{theorem}

\proof

\emph{Membership.}
Given an itemset $P$, a witness to its non maximality/closeness is an itemset 
$Q\supsetneq P$ that  satisfies $C$, and in the case of closeness has 
the same frequency as $P$. 
Checking that $Q$  has the same frequency as $P$
is linear in $||\SDB||$. 
Checking that $Q$ satisfies $C$ requires checking 
the $C$ constraints, which is polynomial in  $|C|\cdot||\SDB||$. 
Hence, the "no" answer admits a polynomial certificate, 
and  deciding maximality or closeness for   $Th_{\SDB}(C)$ is in coNP.  \\
\emph{Completeness.}
We reduce 3-UNSAT, which is coNP-complete,  
to the problem of deciding whether an itemset  $\{\texttt{z}\}$ is maximal/closed. 
Given a 3-CNF formula $F$ with $m$ clauses 
on the set  $V = \{v_1,\ldots, v_n\}$ of 
Boolean variables, we 
want to decide whether $F$  is unsatisfiable. 
A clause $cl_j$ in $F$ is a disjunction 
$(l_{j_1}\lor l_{j_2}\lor l_{j_3})$, where 
a literal $l_i$ denotes either the variable $v_i$ or its negation $\neg v_i$.

We build the dataset  $\SDB$  on the set
$\items=
\{\texttt{pos1}, \ldots, \texttt{posn},  \texttt{neg1}, \ldots,  \texttt{negn}, 
$ $\texttt{cl1},  \ldots, \texttt{clm}, \texttt{z}\}$ of items. 
The intuition is that the pair of items $(\texttt{posi}, \texttt{negi})$  represents 
the Boolean variable $v_i$ in $V$ and the item $\texttt{clj}$  represents 
the clause $cl_j$  in $F$. 
We define the function $item$ such that 
$item(l_i) = \texttt{posi}$ if $l_i=v_i$, and 
$item(l_i) = \texttt{negi}$ if $l_i= \neg v_i$. 
We denote by \texttt{All} the set 
$\{\texttt{pos1}, \texttt{neg1}, \ldots, \texttt{posn}, $ $\texttt{negn}, 
\texttt{cl1},  \ldots,\texttt{clm}, \texttt{z}\}$ of all items. 

The dataset  $\SDB$  contains  $m$ transactions. 
For each  clause $cl_j$ in $F$, $\SDB$ contains the  transaction  
$\texttt{All}\setminus \{\texttt{clj}\}$. 

The set $C$ is composed of the following constraints. 
For each variable  $v_i\in V$, $C$ contains a constraint  $c^{var}_{i}$
defined by
$$\forall P\in 2^{\items},  c^{var}_{i}(P) \equiv
|P\cap \{\texttt{posi},\texttt{negi}\}|\neq 2$$ 
For each clause  $cl_j=(l_{j_1}\lor l_{j_2}\lor l_{j_3})\in F$, 
$C$ contains a constraint  $c^{clause}_{j}$ defined by
$$\forall P\in 2^{\items},  c^{clause}_{j}(P) \equiv$$ $$
P\cap \cup_{ i\in 1..n}\{\texttt{posi},\texttt{negi}\}=\emptyset\ \lor\ 
P\cap \{\texttt{clj},item(l_{j_1}),item(l_{j_2}),item(l_{j_3})\}\neq\emptyset$$
Finally, $C$ contains the frequency constraint $c_{freq}$, with  
the frequency threshold $s$  set to $m$. 

The dataset contains $m$ transactions and the 
set $C$ contains $n+m+1$ constraints. 
The reduction is thus polynomial in size. We now show that deciding 
maximality/closeness of item $\texttt{z}$ is equivalent to deciding 
unsatisfiability of formula $F$.

Suppose the formula $F$ is satisfiable. 
Let us denote by $A$ an assignment satisfying $F$. 
We construct the itemset $P$ containing 
$\texttt{z}$ and 
such that  
for each $i$ such that $A[v_i]=1$, 
$\texttt{posi}\in P$ and $\texttt{negi} \not\in P$, 
and  for each $i$ such that $A[v_i]=0$, 
$\texttt{posi}\not\in P$ and $\texttt{negi} \in P$. 
By  construction of $P$, all constraints $c^{var}$ are satisfied. 
By  construction of $P$, the fact that $A$ satisfies $F$ implies that 
all constraints $c^{clause}$ are satisfied too because for each clause  
$cl_j=(l_{j_1}\lor l_{j_2}\lor l_{j_3})\in F$, at least one literal is true in $A$, 
thus at least one of the items $item(l_{j_1}),item(l_{j_2}),item(l_{j_3})$ 
is in $P$. 
Finally, by  construction of $P$  and $\SDB$, $c_{freq}$ is satisfied 
because $P$ does not contain any item \texttt{clj} and thus all 
the $m$ transactions  contain $P$.  
As a result, $\texttt{z}$ is neither maximal nor closed for   $Th_{\SDB}(C)$ 
because $P$ is a superset of $\texttt{z}$ satisfying $C$ 
and with same frequency $m$ as $\texttt{z}$. 

Suppose now that $\texttt{z}$ is not maximal or not closed for   $Th_{\SDB}(C)$. 
This means that there exists a superset $P$ of $\texttt{z}$ satisfying $C$. 
Thanks to constraint $c_{freq}$ in $C$, 
we know that all $m$ transactions contain $P$. 
Hence, $P$ cannot contain any item $\texttt{clj}$. 
Thus, thanks to constraints $c^{clause}$, we are guaranteed that for 
each clause $cl_j=(l_{j_1}\lor l_{j_2}\lor l_{j_3})\in F$, $P$ contains 
at least one of the items $item(l_{j_1}),item(l_{j_2}),item(l_{j_3})$.  
In addition, thanks to constraints $c^{var}$, we are guaranteed that 
$P$ does not contain both  \texttt{posi} and \texttt{negi} for any variable 
$v_i\in V$. 
As a result, the assignment $A$ built by setting $A[v_i]=1$ when 
$\texttt{posi}\in P$, and $A[v_i]=0$ when $\texttt{posi}\notin P$, 
is a satisfying assignment for $F$.\footnote{Observe that for some $v_i\in V$, 
it is possible that neither $\texttt{posi}$ nor $\texttt{negi}$ are in $P$. 
This can happen when all clauses are satisfied whatever value $v_i$ takes. 
In such a case, we chose to set $v_i$ to 0 in $A$.}

Consequently, 
deciding whether an itemset  $\texttt{z}$ is maximal/closed is equivalent 
to deciding whether a CNF formula $F$ is unsatisfiable, which is coNP-complete. 
\qed

\begin{corollary}
Given a dataset $\SDB$ on a set of items $\items$ and a set $C$ of user's constraints, 
finding an itemset that is maximal/closed
for   $Th_{\SDB}(C)$
is coNP-hard, and it remains so even if $c_{freq}\in C$. 
\end{corollary}

\section{Conclusion}
\label{sec:conclusion}
In this paper we have analyzed the computational complexity of 
some well known itemset mining problems.
We have proved that mining a confident rule that has a given item in the head 
is NP-hard, mining a high utility itemset is NP-hard,  
and mining a maximal or closed constrained itemsets is coNP-hard.
We hope that these results will give directions on which algorithmic technique 
to choose for these problems.

\bibliographystyle{spbasic} 

\bibliography{bib.bib}

\end{document}